\documentstyle[preprint,aps,psfig]{revtex}

\newcommand{\half}{\frac{1}{2}}
\newcommand{\cderiv}[1]{#1 \frac{\partial}{\partial #1}}
\newcommand{\ren}{\mbox{ren}}
\newcommand{\gh}{\mbox{gh}}
\newcommand{\tGamma}{\tilde{\Gamma}}

\begin{document}

\title{Multiplicative BRST renormalization of the $SU(2)$ Higgs model}

\author{Anne Schilling and Peter van Nieuwenhuizen}

\address{Institute for Theoretical Physics, State University
of New York at Stony Brook, Stony Brook, NY 11794-3800}

\preprint{ITP-SB 93-53}
\maketitle
\begin{abstract}
We reformulate the proof of the renormalization of a spontaneously
broken gauge theory by multiplicatively renormalizing the vacuum
expectation value of the Higgs field in the $SU(2)$~Higgs model.
\end{abstract}
\narrowtext

After BRST symmetry was discovered~\cite{BRS1}, it was noted that
the renormalization of gauge theories~\cite{Velt} can be simplified by using
BRST methods. In~\cite{BRS2} Becchi, Rouet and Stora
considered the $SU(2)$ Higgs model, introduced sources for the BRST
variations of the gauge and ghost fields, and made an analysis of the
two-point-functions, the S-matrix, unitarity, and even an cohomologic
analysis of anomalies. However, they did not give a systematic loop by loop
proof of renormalizability. Also, their work was complicated because the
non-nilpotency of the BRST transformations of the antighost introduces ghost
equations of motion at various places. At about the same time,
Zinn-Justin~\cite{Zinn} considered the renormalization of gauge theories loop
by loop
using BRST Ward identities. For spontaneously broken gauge
theories these Ward identities remain valid because they are independent
of the value (in particular the sign) of the mass term in the matter
section. Using this approach, B.Lee~\cite{Lee} gave a systematic analysis
 of the
multiplicative loop-by-loop renormalization of unbroken gauge theories.
This account supercedes his earlier treatments~\cite{LeeZinn} which did not
use BRST
methods and were much less clear. At the end of his analysis he also briefly
considered spontaneously broken gauge theories where he started a proof of the
renormalizability of a general spontaneously broken gauge theory by
first shifting the n-loop renormalized scalars $s_{\alpha}^{\ren}$ over
an amount ${\delta}u_{\alpha}$ such that tadpoles (one particle
irreducible (1PI) 1-point~graphs) were removed.
This is thus an additive renormalization. The remaining parameters he
renormalized multiplicatively. In particular, the scalar fields $s_{\alpha}$
were renormalized as follows
\begin{equation}
s_{\alpha} = Z_{\alpha}^{\half} (s_{\alpha}^{\ren}+u_{\alpha}^{\ren}
             +{\delta}u_{\alpha})
\end{equation}
such that $s_{\alpha}^{\ren}$ vanishes at the minimum of the
renormalized effective potential. In this note we want to demonstrate that
one can
treat all renormalizations on equal footing as multiplicative
renormalizations. In particular, the vacuum expectation value $v$ gets
a $Z$-factor $Z_{v}$ which differs from the $Z$~factor of the corresponding
scalar field.
The advantage of multiplicative instead of additive renormalization is that the
BRST symmetry is manifestly preserved under the renormalization program. If one
only has additive renormalization, one has to prove this property; such a
proof has been
given in reference~\cite{Okawa}.

The model we consider is the $SU(2)$~spontaneously broken gauge theory
coupled to the Higgs sector of the standard model~\cite{Hooft},
with $\sigma$ the Higgs scalar and $\chi^{a}$ the would-be Goldstone
bosons. Surprisingly we find that there is one more divergent structure allowed
by the BRST Ward identities than there are $Z$~factors. This problem is
resolved
because we have found a new identity for the effective action of spontaneously
broken gauge theories, which holds in addition to the BRST Ward identities, and
which originates from the observation that in the matter sector only the
unbroken
$\sigma+v$ appears.

The Lagrangian is given by
\begin{equation}
\label{lagrange}
{\cal L}={\cal L}(\mbox{gauge})+ {\cal L}(\mbox{matter})
          + {\cal L}(\mbox{fix})+ {\cal L}(\mbox{ghost})
          + {\cal L}(\mbox{sources})
\end{equation}
where
\begin{eqnarray}
\label{Lgauge}
{\cal L}(\mbox{gauge})&=&-\frac{1}{4} (\partial_{\mu} A_{\nu}^{a} -
\partial_{\nu} A_{\mu}^{a} + gf^{a}\! _{bc}A_{\mu}^{b}A_{\nu}^{c})^{2}\\
\label{Lmatter}
{\cal L}(\mbox{matter})&=&-\half (D_{\mu}\sigma)^{2} -\half
(D_{\mu}\chi^{a})^{2}
                  +\half \mu^{2} \{(\sigma+v)^{2} + (\chi^{a})^{2}\}
                  -\frac{1}{4} \lambda\{(\sigma+v)^{2} +
(\chi^{a})^{2}\}^{2}\nonumber\\
               &=&-\half (D_{\mu}\sigma)^{2} -\half (D_{\mu}\chi^{a})^{2}
                  -\frac{1}{4} \lambda\{\sigma^{2} + (\chi^{a})^{2}\}^{2}
                  -\lambda v\sigma [\sigma^{2}+(\chi^{a})^{2}]\nonumber\\
                 && -\lambda v^{2}\sigma^{2} - \beta v \sigma
                  -\half \beta (\sigma^{2}+(\chi^{a})^{2})\\
{\cal L}(\mbox{fix})&=&-\frac{1}{2\alpha} (\partial^{\mu}A_{\mu}^{a} - \xi
(\half gv)\chi^{a})^{2}\\
{\cal L}(\mbox{ghost})&=&b_{a}\{\partial^{\mu}D_{\mu}c^{a}
                 -\xi\half gv(\half g(\sigma+v)c^{a} +
\half gf^{a}\! _{bc}\chi^{b}c^{c})\}\\
{\cal L}(\mbox{sources})&=&K^{\mu}_{a}D_{\mu}c^{a} - K(\half g\chi_{a}c^{a})
                   + K_{a}(\half g(\sigma+v)c^{a} +
\half gf^{a}\! _{bc}\chi^{b}c^{c})
                   + L_{a}(\half gf^{a}\! _{bc}c^{b}c^{c})
\end{eqnarray}
The parameter $\beta$ is given by $$\beta = -\mu^{2}+\lambda v^{2}$$
and $-\half\mu^{2}((\sigma+v)^{2}+(\chi^{a})^{2})$ is the mass-term
in the matter section before spontaneous symmetry breaking. Since
$\mu^{2}$ and $\lambda v^{2}$ will in general renormalize differently,
one cannot expect that $\beta$ renormalizes multiplicatively.
It is very convenient to require that the value of the renormalized
$\beta$ be zero
\begin{equation}
\beta_{\ren} = -\mu_{\ren}^{2} + \lambda_{\ren}v_{\ren}^{2}=0
\end{equation}
since this eliminates terms linear in $\sigma$ from the quantum action.
Of course, $\beta_{\ren}=0$ also excludes multiplicative renormalizability
of $\beta$.
In the preceding article~\cite{waldron} we found it useful to consider $\beta$
instead of $\mu^{2}$ as an independent variable, and renormalized
$\beta$ additively. Taking now $\mu^{2}$ as independent variable saves
multiplicative renormalization.

The external sources $K, K_{a}$ and $K^{\mu}_{a}$ multiply the
BRST~variations of $\sigma$,$\chi^{a}$ and $A_{\mu}^{a}$, and the
theory with (and hence without) them will be shown to be renormalizable.
The covariant derivatives are given by
\begin{eqnarray}
D_{\mu}\sigma &=& \partial_{\mu}\sigma + \half gA_{\mu}^{a}\chi_{a}\\
D_{\mu}\chi^{a} &=& \partial_{\mu}\chi^{a}
                  - \half gA_{\mu}^{a}(\sigma+v)
                  + \half gf^{a}\! _{bc}A_{\mu}^{b}\chi^{c}.
\end{eqnarray}

Clearly, ${\cal L}(\mbox{matter})$ depends only on $\sigma+v$, but
${\cal L}(\mbox{fix})$ and ${\cal L}(\mbox{ghost})$ violate this
property for $\xi \neq 0$. Hence we may expect that $\sigma$ and $v$
will renormalize differently if $\xi \neq 0$. We shall assume that the
renormalized $\xi$ and $\alpha$ have 't~Hooft's~\cite{Hooft} values
$\xi_{\ren}=\alpha_{\ren}=1$ in order that the propagators be diagonal
and simple.

The two Ward identities used by B.Lee~\cite{Lee} for the effective action
$\Gamma$ read before renormalization
\begin{eqnarray}
\label{ward1}
&&\partial\tGamma/\partial\Phi^{I} \frac{\partial}{\partial K^{I}}\tGamma =0\\
\label{ward2}
&&(\partial^{\mu} \frac{\partial}{\partial K_{a}^{\mu}}
 -\xi\half gv\frac{\partial}{\partial K_{a}}
 -\frac{\partial}{\partial b_{a}}) \tGamma = 0
\end{eqnarray}
where $\Phi^{I} = \{\sigma,\chi^{a},A_{\mu}^{a},c^{a}\},
K^{I} = \{K,K_{a},K^{\mu}_{a},L_{a}\}$ and
$\tGamma = \Gamma - \int{\cal L}(\mbox{fix})d^{4}x$.
In addition we shall use below two further identities related to ghost
number conservation and to the symmetry of ${\cal L}(\mbox{matter})$
under $\sigma \rightarrow \sigma+\Delta v, v \rightarrow v-\Delta v$.

The Ward~identities in~(\ref{ward1}) and~(\ref{ward2}) remain valid after
renormalization if all rescalings are such that they amount to an overall
factor. Choosing $A_{\mu}^{a}=(Z_{3})^{\half}A_{\mu,\ren}^{a}$
and $c^{a}=(Z_{\gh})^{\half}c^{a}_{\ren}$, and furthermore
$\sigma=(Z_{\sigma})^{\half}\sigma_{\ren}$ and
$\chi^{a}=(Z_{\chi})^{\half}\chi^{a}_{\ren}$ ($Z_{\chi}$ is independent of the
$SU(2)$~index $a$ since ${\cal L}(\mbox{fix})$ is $SU(2)$~invariant),
we assume,
to be proven by induction in order of loops, the following properties:
\begin{enumerate}
\item $\tGamma$ is made finite by multiplicative rescalings of all objects.
 In particular,
    $K^{\mu}_{a}$ and $b_{a}$ scale like $c^{a}$, while $L_{a}$
    scales like $A_{\mu}^{a}$. Furthermore the scales of $K$ and $K_{a}$
    are such that $\sigma K$ and $\chi^{a}K_{a}$ have the same $Z$~factor
    as $A_{\mu}^{a}K_{a}^{\mu}$.
\item $\alpha$ and $\xi$ must scale such that ${\cal L}(\mbox{fix})$
    is finite by itself since we now deal with $\tGamma$ which is
    $\Gamma$ minus $\int{\cal L}(\mbox{fix})d^{4}x$.
\end{enumerate}

We also renormalize $g=Z_{g}g_{\ren}, v=(Z_{v})^{\half}v_{\ren},
\lambda=Z_{\lambda}\lambda_{\ren}$ and $\mu^{2}=Z_{\mu^{2}}\mu^{2}_{\ren}$.
Hence
\begin{eqnarray}
K&=&(Z_{3}Z_{\gh}/Z_{\sigma})^{\half}K_{\ren}\nonumber\\
K^{a}&=&(Z_{3}Z_{\gh}/Z_{\chi})^{\half}K^{a}_{\ren}\nonumber\\
\label{Zfactors}
\xi&=&Z_{3}^{\half}Z_{g}^{-1}Z_{v}^{-\half}Z_{\chi}^{-\half}\xi_{\ren}
\end{eqnarray}

The equality of the $Z$-factors of $b$ and $c$ is not a matter of choice
because in ${\cal L}(\mbox{source})$ there are terms with $c$ but without
$b$.

\begin{figure}
\vspace{.5cm}
\hbox{
	\psfig{figure=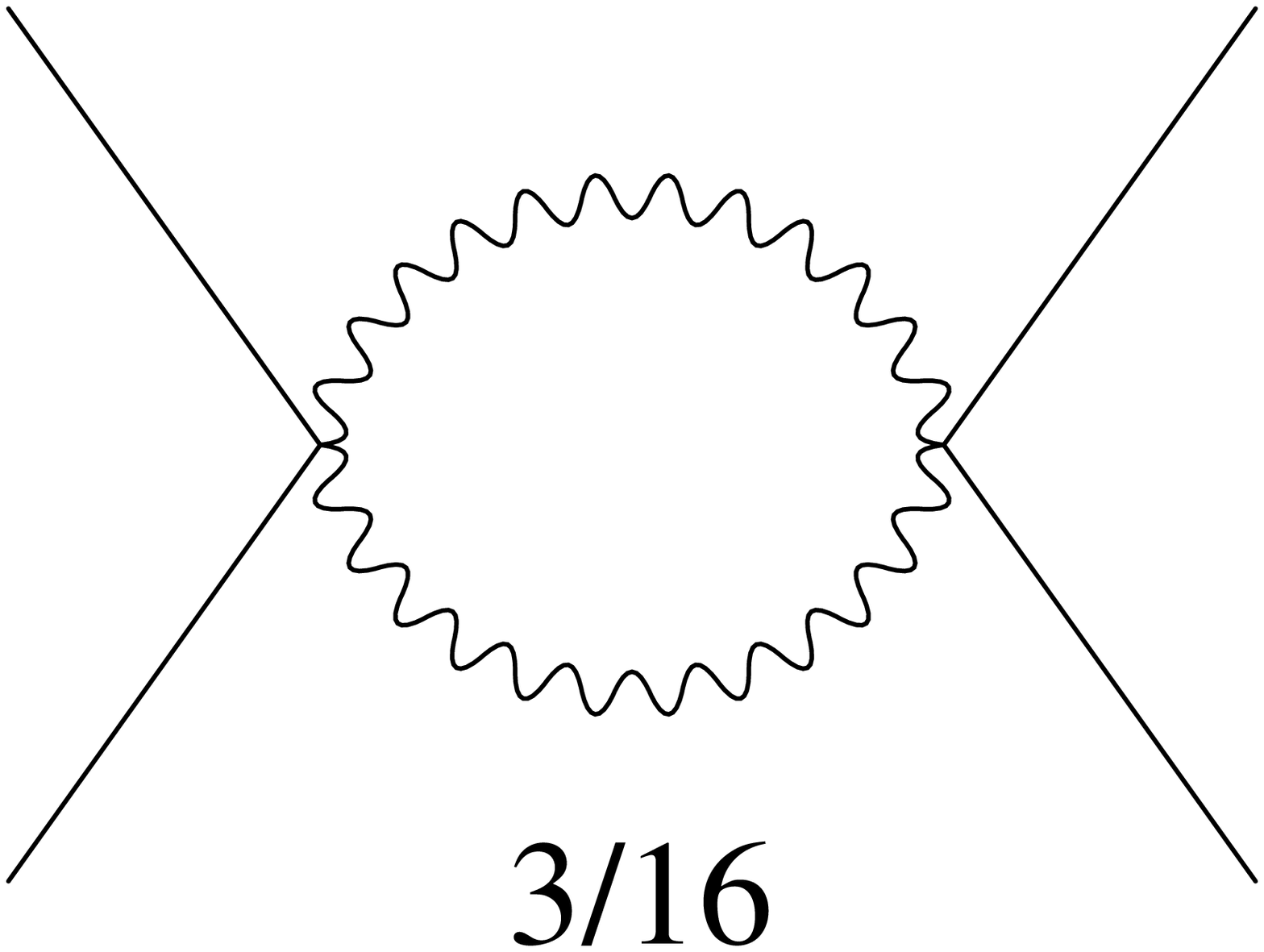,height=2.3cm}
	\hspace{.3cm}
	\psfig{figure=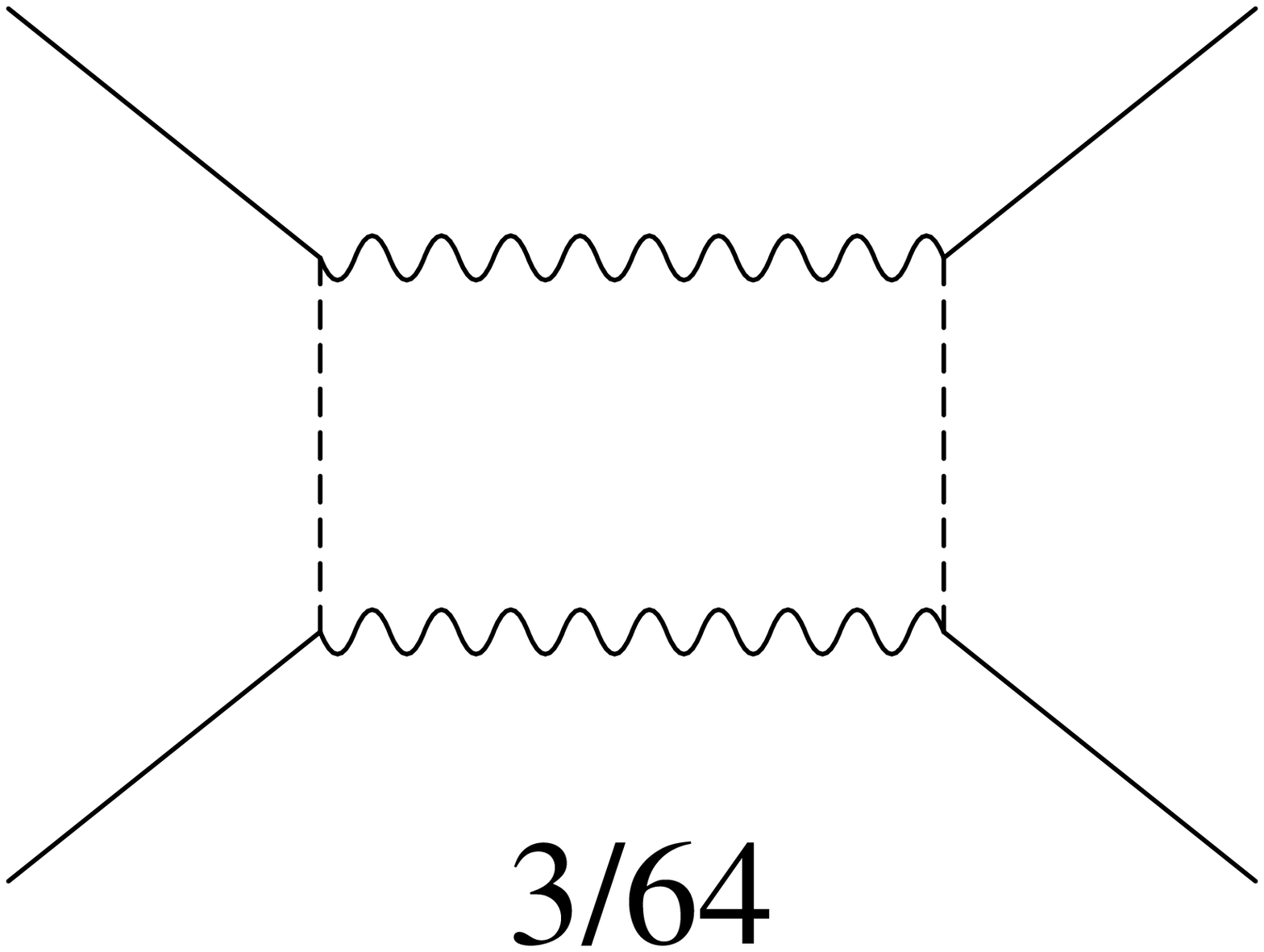,height=2.3cm}
	\hspace{.3cm}
	\psfig{figure=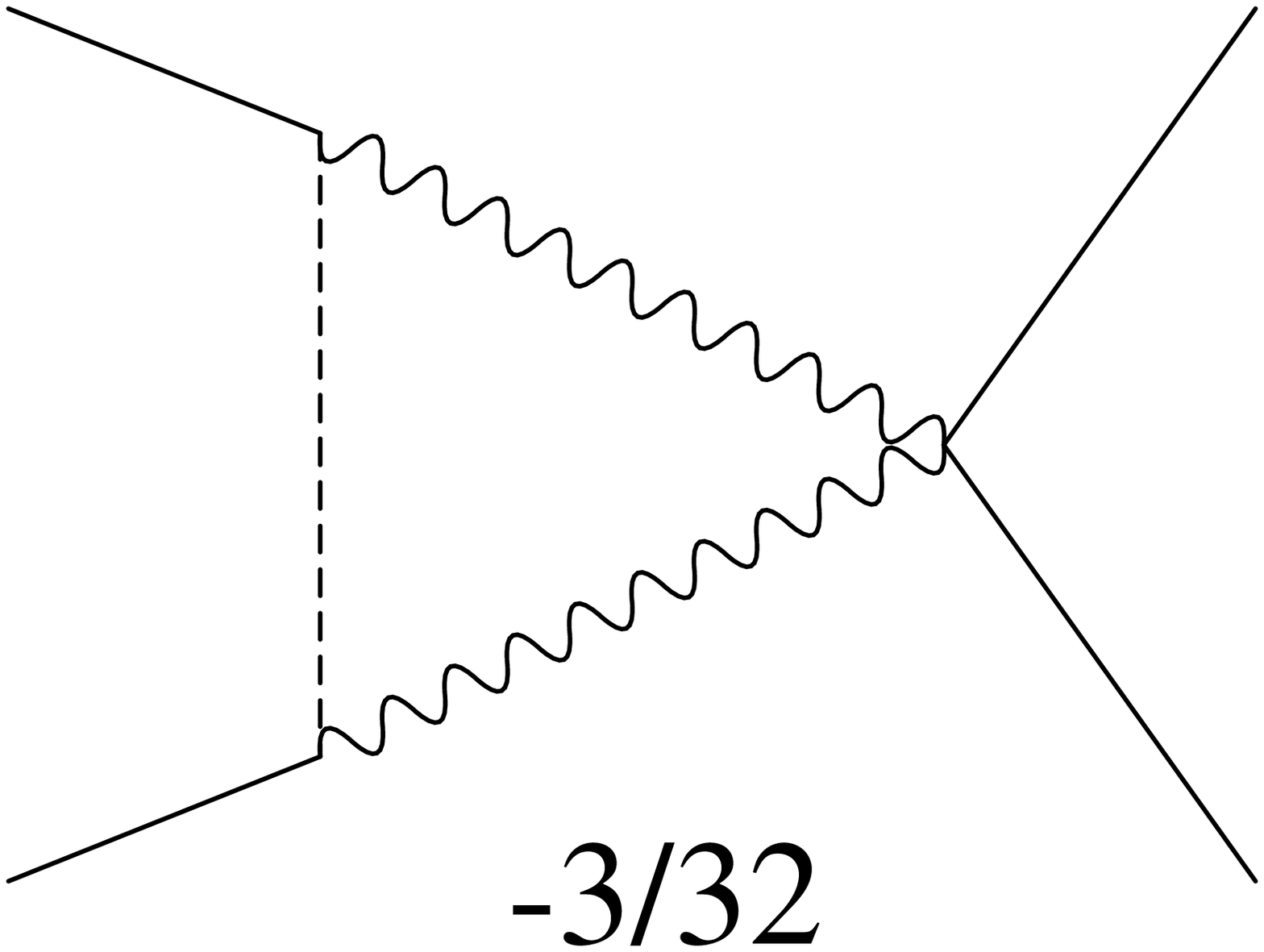,height=2.3cm}
	\hspace{.3cm}
	\psfig{figure=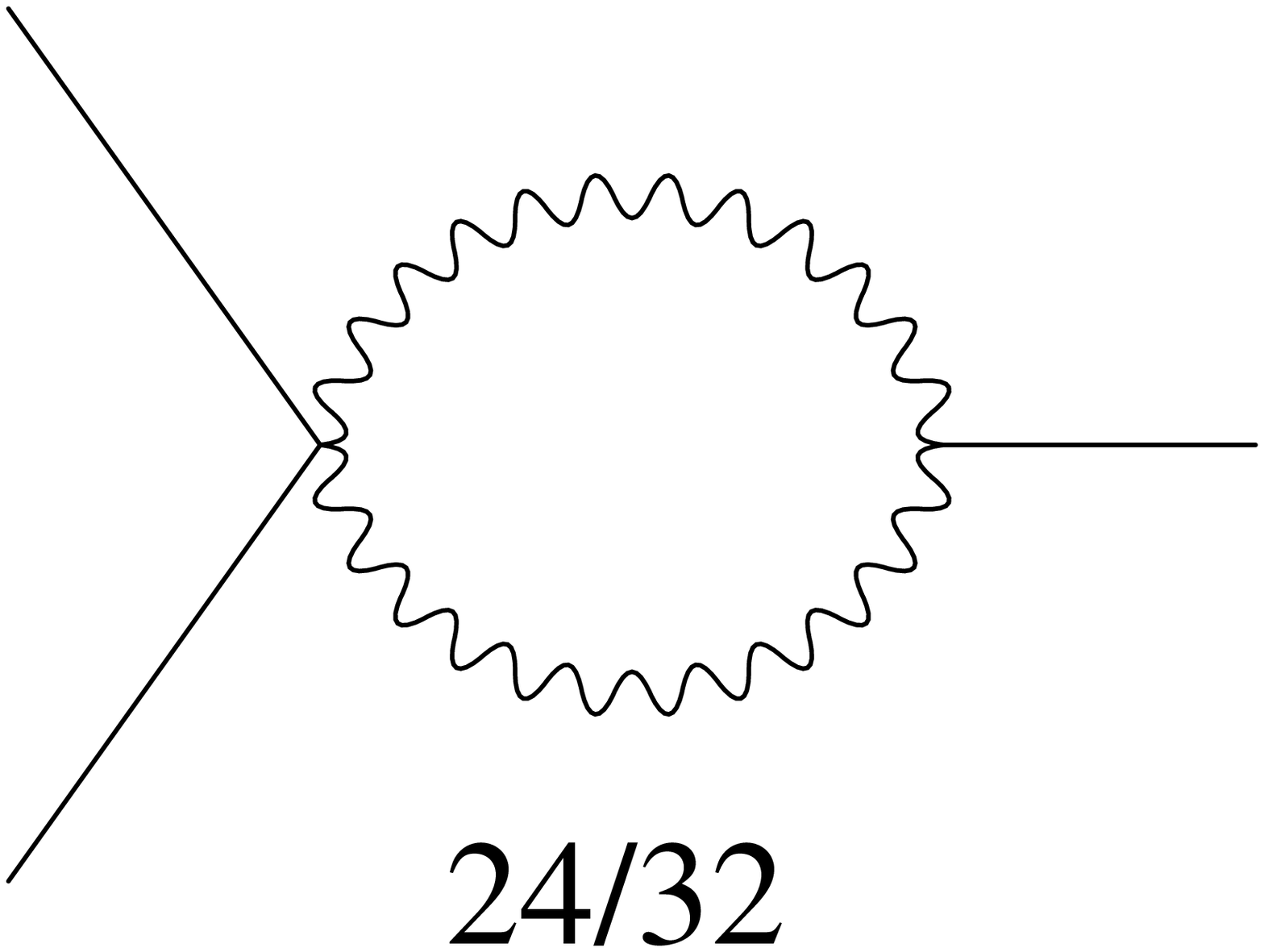,height=2.3cm}
	\hspace{.3cm}
	\psfig{figure=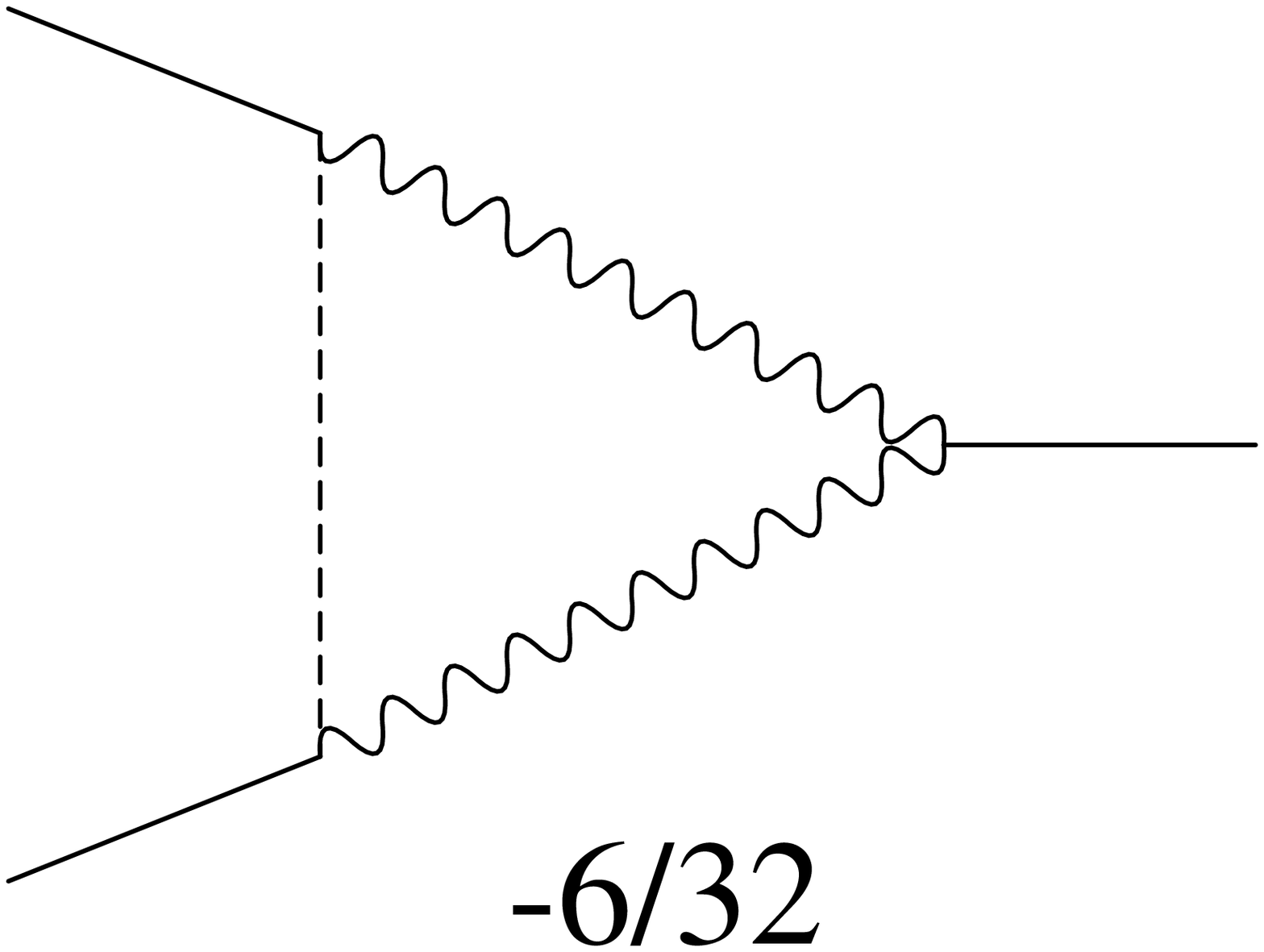,height=2.3cm}
}
\vspace{.7cm}

Wiggly lines denote gauge fields, dotted lines denote would-be Goldstone bosons
\caption{\label{figure} The $\sigma^{4}$ and $\sigma^{3}$ one-loop vertex
corrections}
\end{figure}

It is instructive to do a quick one-loop analysis of the $\sigma^{4}$
and $\sigma^{3}$ 1PI~Green's functions to convince oneself that $Z_{v}$
is not equal to $Z_{\sigma}$. In figure~\ref{figure} we have given the
coefficients of the divergences of the relevant divergent graphs.
Clearly four times the sum of the first three coefficients does not equal
the sum of the last two coefficients which shows that $Z_{v}\neq Z_{\sigma}$
in the gauge sector. Since in the matter sector
$Z_{v}=Z_{\sigma}$ (the $\chi$~mass from ${\cal L}(\mbox{fix})$ does
not change this result since massless tadpoles  cancel each other
without having to assume that $\int \frac{d^{4}k}{k^{2}}=0$,
see~\cite{waldron}), we
see that in the model given by~(\ref{lagrange}) one has $Z_{v}\neq Z_{\sigma}$.
We expect $Z_{v}= Z_{\sigma}$ if $\xi=0$. Computations with $\xi=0$ are
somewhat
complicated because propagators are off-diagonal.
The $Z$-factors for the $\sigma$ and
$\chi^{a}$ fields are the same because the terms in the action of dimension
4 have a $SO(4)$ symmetry, even after gauge fixing. (In the notation of
reference~\cite{Hooft}, the $SU(2)$ gauge transformation acts from the left on
$D_{\mu}(\sigma+i\chi^{a}\tau_{a})$, while the other $SU(2)$ group is rigid,
acting from the right leaving $A_{\mu}^{a}$, $c^{a}$ and $b_{a}$ invariant).

Assuming $(n-1)$-loop finiteness of $\tGamma$ (and hence of $\Gamma$),
the $n$-loop 1PI~divergences satisfy the equations

\begin{equation}
\label{div1}
Q_{\ren}\tGamma_{\ren}^{\mbox{(n),div}}=0
\end{equation}
where
$Q_{\ren}=\partial\tGamma_{\ren}^{(0)}/\partial\Phi_{\ren}^{I}
  \frac{\partial}{\partial K^{\ren}_{I}}
  -\partial\tGamma_{\ren}^{(0)}/\partial K^{\ren}_{I}
  \frac{\partial}{\partial\Phi_{\ren}^{I}}$
and

\begin{equation}
\label{div2}
(\partial^{\mu}\frac{\partial}{\partial K_{a,\ren}^{\mu}}
   - \xi_{\ren}(\half gv)_{\ren}\frac{\partial}{\partial K_{a}^{\ren}}
   - \frac{\partial}{\partial b_{a}^{\ren}})\tGamma_{\ren}^{\mbox{(n),div}}=0
\end{equation}
where $\tGamma_{\ren}^{(0)}$ equals the quantum action minus
$\int{\cal L}(\mbox{fix})d^{4}x$, all in terms of objects
multiplicatively renormalized such that all 1PI~graphs with $(n-1)$~loops
are finite. We shall drop the subscripts ``ren'', understanding that
from now on all objects are $(n-1)$~loop renormalized.

The $n$-loop divergences are local, and~(\ref{div2}) states that $b_{a}$ can
only appear in the divergences in the combination
$K_{a}^{\mu}-\partial^{\mu}b_{a}$ or $K_{a}-\xi\half gvb_{a}$.
This excludes divergences proportional to $b_{a}\partial^{\mu} A_{\mu}^{a}$,
$vb_{a}\chi^{a}$ or $b_{a}\sigma \chi^{a}$.
The general form of the $n$-loop divergences is given by
\begin{equation}
\label{sol}
\tGamma_{\ren}^{\mbox{(n),div}}=
      \sum_{i=1}^{4} a_{i}(\epsilon)G^{i}
      +Q_{\ren}\sum_{j=1}^{5}b_{j}(\epsilon)X^{j}
\end{equation}
where the first term contains all possible gauge-invariant local expressions,
see~(\ref{Lgauge}) and~(\ref{Lmatter}),
\begin{equation}
\sum a_{i}G^{i}=a_{1}S(\mbox{gauge})+a_{2}S(\mbox{kin. matter})
                +a_{3}S(\mbox{mass matter})+a_{4}S(\mbox{pot.})
\end{equation}
while the second term is given by
\begin{equation}
\sum b_{j}X^{j}= b_{1}(K_{a}^{\mu}-\partial^{\mu}b_{a})A_{\mu}^{a}
               + b_{2}(K_{a}-\half gv\xi b_{a})\chi^{a}
               + b_{3}K\sigma
               + b_{4}L_{a}c^{a}
               + b_{5}Kv.
\end{equation}

Because of the $SO(4)$ symmetry, $b_{2}=b_{3}$, but we shall keep writing
$b_{2}$ and $b_{3}$ separately in order to facilitate the identification
of divergences.
It is easy to see that~(\ref{sol}) is a solution of~(\ref{div1})
and~(\ref{div2}) since $Q$, the BRST~charge, acting on a gauge invariant
term is zero and $Q^{2}=0$ (see~\cite{Lee} and~\cite{Itzykson}).
In~\cite{Joglekar} a general
(model independent) but rather complicated (and incomplete) proof is
given that the general
solution of~(\ref{div1}) is a sum of gauge invariant terms
and $Q$-exact terms as in~(\ref{sol}).
It is possible to prove this for a given model in a simple
and direct way as follows:

\begin{enumerate}
\item write down all local expressions with dimension four and
ghost-number zero which can be a priori divergent according to power
counting
\item use the fact that their sum must be annihilated by
$Q_{\ren}$.
\end{enumerate}

For the model in~(\ref{lagrange}), the result is~(\ref{sol}).

We observe that there are eight divergent structures but only seven $Z$-factors
(for $A_{\mu}^{a}$,$\sigma$ and $\chi^{a},c^{a},g,v,\lambda,\mu^{2}$). In pure
unbroken Yang-Mills theory there is no such mismatch, but in the matter
coupled case with unbroken symmetry the same mismatch occurs. As we shall see,
multiplicative renormalizability is still possible because the eight
divergences $a_{i}$ and $b_{j}$ (where $b_{2}=b_{3}$) only occur in seven
combinations.

To prove multiplicative renormalizability, each of the local
divergences should be written as a counting operator
$\cderiv{x}$ acting on $\tGamma^{(0),\ren}$ where
x denotes all fields, sources and parameters in the theory. For most
terms, the analysis has already been given by B.Lee~\cite{Lee}.
In particular
\begin{eqnarray}
S(\mbox{gauge})&=&\frac{1}{g^{2}}S(gA_{\mu}^{a})\\
               &=&(-\half \cderiv{g}
                 +\half \cderiv{A_{\mu}^{a}}
                 +\half \cderiv{L_{a}}\nonumber\\
               &&+\half \cderiv{K}
                 +\half \cderiv{K_{a}}
                 +\cderiv{\xi})\tGamma_{\ren}^{(0)}\\
S(\mbox{kin. matter})&=&(\half \cderiv{\sigma}
                 +\half \cderiv{v}
                 +\half \cderiv{\chi_{a}}
                 -\cderiv{\mu^{2}}\nonumber\\
               &&-2\cderiv{\lambda}
                 -\half \cderiv{K}
                 -\half \cderiv{K_{a}}
                 -\cderiv{\xi})
                      \tGamma_{\ren}^{(0)}.
\end{eqnarray}

The terms from $QX$ lead to the counting operators
\begin{eqnarray}
b_{1}(\cderiv{A_{\mu}^{a}}-(K_{a}^{\mu}-\partial^{\mu}b_{a})\frac{\partial}{\partial K_{a}^{\mu}})
+b_{2}(\cderiv{\chi_{a}}-(K_{a}-\half gv\xi b_{a})\frac{\partial}{\partial
K_{a}})\nonumber\\
+b_{3}(\cderiv{\sigma}-\cderiv{K})
+b_{4}(\cderiv{c_{a}}-\cderiv{L_{a}})
+b_{5}v \frac{\partial}{\partial\sigma}.
\end{eqnarray}

Most terms in $QX$ are already of the form
$\cderiv{x}\tGamma_{\ren}^{(0)}$. We now analyze the terms which are not
yet cast into
this form
\begin{equation}
(b_{1}\partial^{\mu}b_{a} \frac{\partial}{\partial K^{\mu}_{a}}
+b_{2}\xi\half gvb_{a} \frac{\partial}{\partial K_{a}}
+b_{5}v \frac{\partial}{\partial\sigma})\tGamma^{(0)}_{\ren}.
\end{equation}

The first term equals $-S(\mbox{ghost})$ at $\xi=0$, and can be written
as
$-\half\cderiv{b_{a}}-\half\cderiv{c_{a}}+\half\cderiv{K}+\half\cderiv{K_{a}}
+\half\cderiv{K_{a}^{\mu}}+\cderiv{L_{a}}+\cderiv{\xi}$ acting on
$\tGamma^{(0)}_{\ren}$.
The second term is minus the $\xi$ term in $S(\mbox{ghost})$, hence it
equals $-\cderiv{\xi}$ acting on $\tGamma^{(0)}_{\ren}$. The last term
we deal with later.

Analyzing these results, we see that the combination
$\cderiv{c_{a}}+\cderiv{b_{a}}$ appears
everywhere except in the term with $b_{4}$.
However, ghost number conservation leads to the Ward~identity
\begin{equation}
(\cderiv{b_{a}}-\cderiv{c_{a}}+\cderiv{K}+\cderiv{K_{a}}+\cderiv{K_{a}^{\mu}}+2\cderiv{L_{a}})
\tGamma^{(0)}_{\ren}=0
\end{equation}
and using this identity to convert half of the $b_{4}$~terms, we find
also in the $b_{4}$~term the desired combination
$\cderiv{c_{a}}+\cderiv{b_{a}}$. At this point the divergences can be
written as
\begin{eqnarray}
\tGamma^{\mbox{(n),div}}_{\ren}=&[&
   (\half a_{1}+b_{1})(\cderiv{A_{\mu}^{a}}+\cderiv{L_{a}})\nonumber\\
&+&(-\half b_{1}+\half
b_{4})(\cderiv{c_{a}}+\cderiv{b_{a}}+\cderiv{K_{a}^{\mu}})\nonumber\\
&+&(\half a_{2}+b_{3})\cderiv{\sigma}+(\half
a_{2}+b_{2})\cderiv{\chi_{a}}\nonumber\\
&+&(\half a_{1}-\half a_{2}+\half b_{1}-b_{3}+\half b_{4})\cderiv{K}\nonumber\\
&+&(\half a_{1}-\half a_{2}+\half b_{1}-b_{2}+\half
b_{4})\cderiv{K_{a}}\nonumber\\
&-&\half a_{1}\cderiv{g}+(a_{1}-a_{2}+b_{1}-b_{2})\cderiv{\xi}\nonumber\\
&+&(-a_{2}+a_{3})\cderiv{\mu^{2}}+(-2a_{2}+a_{4})\cderiv{\lambda}\nonumber\\
&+&\half a_{2}\cderiv{v}+b_{5}v\frac{\partial}{\partial\sigma}
]\tGamma^{(0)}_{\ren}.
\label{counting}
\end{eqnarray}

Since $b_{2}=b_{3}$, we see that indeed $Z_{\sigma}=Z_{\chi}$ and the
$Z$-factors for $K$ and $K_{a}$ are equal. We also
see that $A_{\mu}^{a}$ and $L_{a}$ scale the same way as do
$c^{a},b_{a}$ and $K_{a}^{\mu}$. Furthermore, the factors in front of
$\cderiv{K}$ (or$\cderiv{K_{a}}$) depend linearly on those corresponding
to $A_{\mu}^{a}, c^{a}$ and $\sigma$ (or $\chi^{a}$), namely in
agreement with~(\ref{Zfactors}).

As usual, $\alpha=Z_{3}\alpha_{\ren}$ fulfills step 2 of the
induction as far as the $\xi$-independent terms are concerned. We are left
with the only nontrivial part of the proof of renormalizability,
the proof that the
rescaling of $\xi$ in~(\ref{Zfactors}) is consistent with the rescaling
of $v$ which has
been left unspecified so far. The key to this compatibility lies in the
last term
in~(\ref{counting}), the term with
$b_{5}v\frac{\partial}{\partial\sigma}\tGamma^{(0)}_{\ren}$.
To write it down, too, as a counting operator, we recall that the matter action
is annihilated by $v\frac{\partial}{\partial\sigma}-\cderiv{v}$ (since
it only depends on $\sigma+v$).
There is only one place in $\tGamma_{\ren}^{(0)}$ where $v$ appears
separately, namely in the term
with $\xi$ in ${\cal L}(\mbox{ghost})$. (Recall that in
$\tGamma$ there is no ${\cal L}(\mbox{fix})$). Clearly then, the
following identity holds
\begin{equation}
(\cderiv{\xi}-\cderiv{v}+v\frac{\partial}{\partial\sigma})\tGamma^{(0)}_{\ren}=0.
\end{equation}

Using this identity to eliminate $v\frac{\partial}{\partial\sigma}$
from~(\ref{counting}),
we find that $v$ rescales with $\half a_{2}+b_{5}$
and $\xi$ with $(a_{1}-a_{2}+b_{1}-b_{2}-b_{5})$. These renormalizations
of $\xi$ and $v$ are then also in agreement with the rescaling of $\xi$
in~(\ref{Zfactors}). Hence the multiplicative renormalizability of the
spontaneously
broken $SU(2)$~Higgs~model is proven.

\acknowledgements
We thank John Collins for discussions. Further, we would like to thank
Andrew Waldron
for discussions.
This research was supported in part by the National Science Foundation
under grant
PHY-92-11367.

\begin{references}
\bibitem{BRS1}C. Becchi, A. Rouet, R. Stora, Phys. Lett. {\bf 52B} (1974),
344\\
	      C. Becchi, A. Rouet, R. Stora, Com. Math. Phys. {\bf 42}, 127 (1975)\\
	      I. V. Tyutin, Lebedev preprint, FIAN no 39 (1975), in Russian,
unpublished
\bibitem{Velt}G. 't Hooft, Nucl. Phys. B {\bf33}, 173 (1971) and Nucl. Phys. B
{\bf 35}, 167 (1971)\\
G. t' Hooft and M. Veltman, Nucl. Phys. B {\bf 44}, 189 (1972)
\bibitem{BRS2}C. Becchi, A. Rouet, R. Stora, Ann. Phys. {\bf 98}, 287 (1976)
\bibitem{Zinn}J. Zinn-Justin, in ''Trends in Elementrary Particle
            Theory'', Lecture notes in physics vol 37, Springer Verlag, Berlin,
1975, eds. H.Rollnik and K.Dietz
\bibitem{Lee}B. Lee, in ''Methods in Field Theory, proceedings les Houches
            session 28'', 1975, eds. R.Balian and J.Zinn-Justin,
            North-Holland, Amsterdam, 1976
\bibitem{LeeZinn}B. Lee and J. Zinn-Justin, Phys. Rev. D {\bf 5}, 3137 (1972)
and Phys. Rev. D {\bf 7}, 1049 (1973).
          In these papers, no BRST methods are used.
\bibitem{Okawa}M. Okawa, Progr. of Theo. Phys. {\bf 60}, 1175 (1978)
\bibitem{Hooft}G. 't Hooft, Nucl. Phys.B {\bf 33}, 173 (1971) and {\bf 35}, 167
(1971)
\bibitem{waldron}A. Waldron and P. van Nieuwenhuizen, preprint ITP-SB 93-52
\bibitem{Itzykson}C. Itzykson and J.B. Zuber, Quantum Field Theory,
            McGraw-Hill Book Company, 1980
\bibitem{Joglekar}B. Lee and S.D. Joglekar, Annals Phys. {\bf 97}, 160 (1976)
\bibitem{Dixon}J.A. Dixon, Nucl.Phys.B {\bf 99}, 420 (1975)
\end{references}
\end{document}